\documentclass[aps,showpacs,twocolumn]{revtex4}
\usepackage{amssymb}
\usepackage{amsmath}
\usepackage{graphicx}
\usepackage{epsfig}

\begin{document}

\title{Physics counterpart of the $\mathcal{PT}$ non-hermitian tight-binding
chain}
\author{L. Jin and Z. Song}
\email{songtc@nankai.edu.cn}
\affiliation{School of Physics, Nankai University, Tianjin 300071, China}

\begin{abstract}
We explore an alternative way of finding the link between a $\mathcal{PT}$\
non-Hermitian Hamiltonian and a Hermitian one. Based on the analysis of the
scattering problem for an imaginary potential and its time reversal process,
it is shown that any real-energy eigenstate of a $\mathcal{PT}$
tight-binding lattice with on-site imaginary potentials shares the same wave
function with a resonant transmission state of the corresponding Hermitian
lattice embedded in a chain. It indicates that the $\mathcal{PT}$ eigenstate
of a $\mathcal{PT}$\ non-Hermitian Hamiltonian\ has connection to the
resonance transmission state of the extended Hermitian Hamiltonian.
\end{abstract}

\maketitle

%03.65.-w Quantum mechanics
%11.30.Er Charge conjugation, parity, time reversal, and other discrete symmetries
%71.10.Fd Lattice fermion models (Hubbard model, etc.)
%73.22.Dj Single particle states (ignored)
%73.23.-b Electronic transport in mesoscopic systems (ignored)

\section{Introduction}

Imaginary potential usually appears in a system to describe physical
processes phenomenologically due to its simplicity, which has been
investigated under the non-Hermitian quantum mechanics framework \cite%
{Klaiman1,Znojil,Makris,Musslimani,Bender
08,Jentschura,Fan,A.M38,A.M391,A.M392}. To discuss and explore the
usefulness of the imaginary potential, one has to be able to establish a
correspondence between a non-Hermitian system and a real physical system in
an analytically exact manner. The effort to establish a parity-time ($%
\mathcal{PT}$) symmetric quantum theory as a complex extension of the
conventional quantum mechanics \cite{A.M,A.M36,Jones,Bender 99,Dorey
01,Bender 02,A.M43} was stimulated by the discovery that a non-Hermitian
Hamiltonian with simultaneous $\mathcal{PT}$ symmetry\ has an entirely real
quantum-mechanical energy spectrum \cite{Bender 98} and has profound
theoretical and methodological implications.

When speaking of physical significance of a non-Hermitian Hamiltonian, it is
implicitly assumed that there exists another Hermitian Hamiltonian which
shares the complete or partial spectrum with it. Thus one of the ways of
extracting the physical meaning of a pseudo-Hermitian Hamiltonian having a
real spectrum is to seek for its Hermitian counterparts \cite%
{A.M38,A.M391,A.M392}. The metric-operator theory outlined in \cite{A.M}
provides a mapping of such a pseudo-Hermitian Hamiltonian to an equivalent
Hermitian Hamiltonian. However, the obtained equivalent Hermitian
Hamiltonian is usually quite complicated \cite{A.M,JLPT} and cannot be
judged whether it describes real physics or is just unrealistic mathematical
object.

In this paper, we try to find an alternative way to establish the connection
between a pseudo-Hermitian Hamiltonian and a physics system. We consider a
simple class of discrete systems, which are originally exploited to describe
the solid-state system in condensed matter physics. In such systems, the
imaginary potential usually appears as source or sink, acting as connection
to the outer world. In this sense, the eigenstates of an unbroken $\mathcal{%
PT}$\ non-Hermitian Hamiltonian seem to be the dynamical equilibrium states
of an open system. The strategy of this paper is to seek the ways of
analytical continuation of the eigenfunctions of a $\mathcal{PT}$\textbf{\ }%
non-Hermitian Hamiltonian into the stationary scattering states of an
extended Hermitian system. It indicates that the $\mathcal{PT}$ eigenstate
of a $\mathcal{PT}$\ non-Hermitian Hamiltonian\ has connection to the
resonance transmission state of the extended Hermitian Hamiltonian.

This paper is organized as follows, in Sec. II, Scattering problem of
imaginary potential. In Sec. III, the connection between a resonant
transmission state and a $\mathcal{PT}$ symmetry eigenstate is established.
In Sec. IV, the illustrative examples are presented to demonstrate the main
idea of this paper. Sec. V is the summary and discussion.

\section{Scattering problem of imaginary potential}

The first problem we address in our search for physically meaningful $%
\mathcal{PT}$\textbf{\ }non-Hermitian Hamiltonian is how to associate the
individual imaginary potential with the Hermitian sub-network. Recently the
formal theory of scattering for complex potentials in one dimension
continuous system has been constructed (for review see Ref. \cite{Muga} and
references therein). To establish such a correspondence in a discrete
system, we consider a simple model described by a non-Hermitian Hamiltonian $%
H_{\gamma }$. It is a tight-binding chain with uniform nearest neighbor
hopping integrals and an additional imaginary on-site potential on a site of
a semi-infinite chain, which can be written as follows:
\begin{eqnarray}
H_{\gamma } &=&H_{l\gamma }+H_{g}+H_{\text{sub}},  \label{H_g} \\
H_{l\gamma } &=&-J\overset{-1}{\sum_{l=-\infty }}\left( a_{l}^{\dag }a_{l+1}+%
\text{H.c.}\right) -i\gamma a_{0}^{\dag }a_{0},  \notag \\
H_{g} &=&-g\left( a_{0}^{\dag }a_{1}+a_{1}^{\dag }a_{0}\right) ,  \notag \\
H_{\text{sub}} &=&\sum_{i,j=1}^{N_{s}}\kappa _{ij}\left( a_{i}^{\dag }a_{j}+%
\text{H.c.}\right) ,  \notag
\end{eqnarray}%
where $a_{l}^{\dag }$ is the creation operator of the boson (or fermion) at
the $l$th site, the tunneling strengths and imaginary potential are denoted
by $J$, $g$,\ and $-i\gamma $ $\left( \gamma >0\right) $. We separate the
Hamiltonian into severval parts to characterize the configuration of the
system. A sketch of such a system is given in Fig. 1(a).\ Here $H_{l\gamma }$
is a semi-infinite uniform chain with one potential $-i\gamma $ at the edge,
$H_{g}$ represents the coupling between this chain and an arbitrary
sub-network described by a Hermitian Hamiltonian $H_{\text{sub}}$. In this
sense, the conclusion will be obtained is applicable for a large class of
systems.

\bigskip
%%%%%%%%%%%%%%%%%%%%%%%%%%%%%%%%%%%%%%%%%%%%%%%%%%%%%%%%%%%%%%%%%%%%%%%%
\begin{figure}[tbp]
\includegraphics[ bb=85 96 485 695, width=3.8 cm, clip]{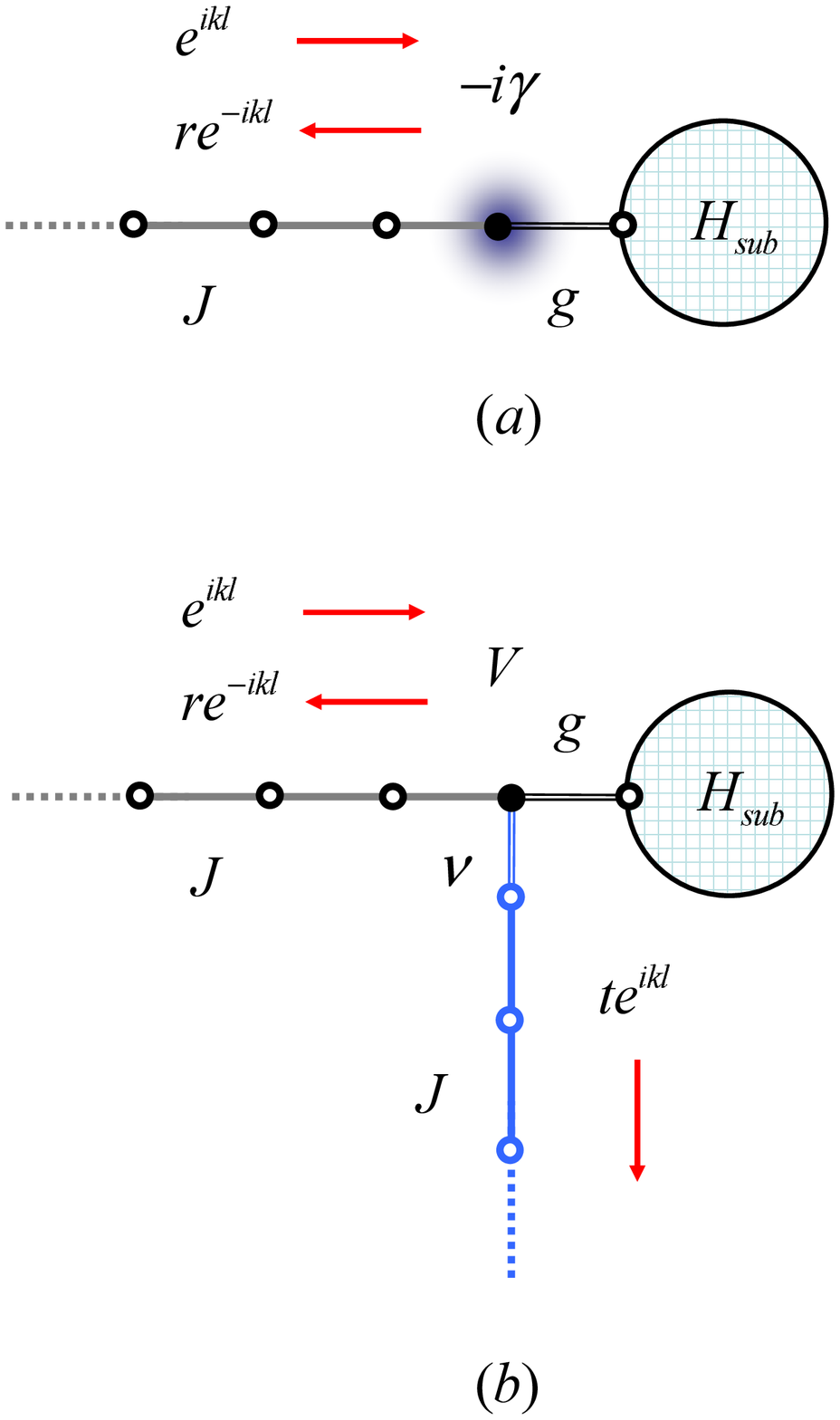}%
\includegraphics[ bb=65 96 465 695, width=3.8 cm, clip]{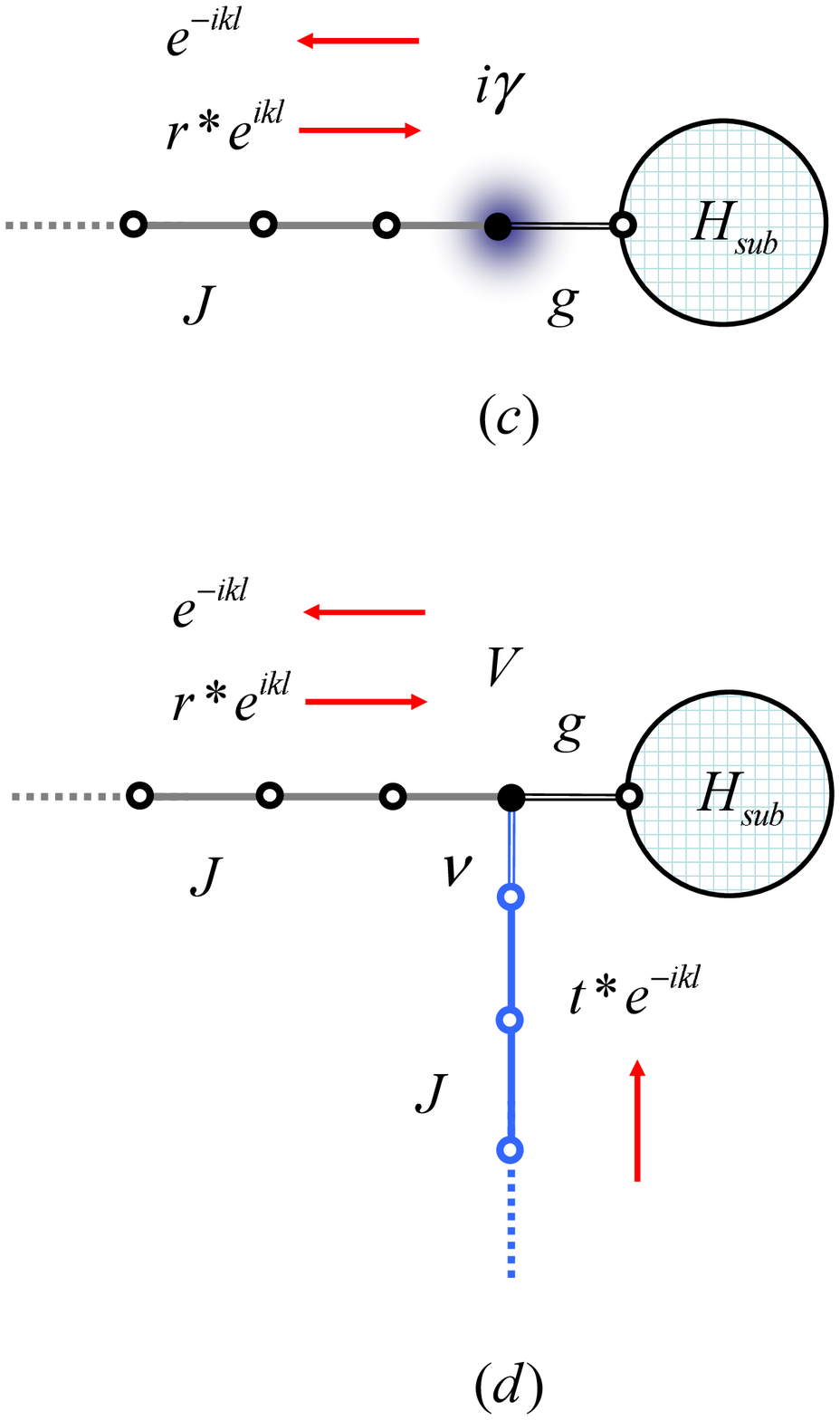}
\caption{(Color online) Schematic illustration of configurations of the
typical tight-binging networks with imaginary potential and their Hermitian
counterparts. (a) the scattering process of an absorptive (absorbing)
imaginary on-site potential, where $H_{\text{sub}}$\ is an arbitrary
Hermitian sub-network. (b) the corresponding Hermitian counterpart network
of (a), which ensures the same wave function as (a) within the common
region. The system consists of system (a) and an attaching lead. (c) and (d)
describe the situations of (a) and (b) under the time reversal operations.
Note that (c) represents the scattering process of a new network different
from (a), while (d) corresponds to the time reversal scattering process of
(c) in the same system.}
\end{figure}

%%%%%%%%%%%%%%%%%%%%%%%%%%%%%%%%%%%%%%%%%%%%%%%%%%%%%%%%%%%%%%%%%%%%%%%%

To investigate the role of the imaginary potential in a discrete system, we
will be concerned with the scattering problem of such system: an incident
plane wave $e^{ikj}$\ or a broad wave packet\ comes from leftmost and is
reflected and transmitted at the imaginary potential. The process can be
represented by the wave function $f_{1}\left( j\right) a_{j}^{\dag
}\left\vert 0\right\rangle $\ $\left( j\in (-\infty ,N_{s}]\right) $\ with
\begin{equation}
f_{1}\left( j\leq 0\right) =e^{ikj}+r_{1}e^{-ikj},  \label{f1}
\end{equation}%
where $r_{1}$\ represents the reflection amplitude. The explicit form of
wave function $f_{1}\left( j>0\right) $ within the sub-network depends on
the the structure of $H_{\text{sub}}$.\ General speaking, the solution of $%
r_{1}$ can not be obtained exactly even the explicit form of $H_{\text{sub}}$%
\ is given. However, we will see that it does not affect the conclusion
below.

In the basis $\left\{ a_{j}^{\dag }\left\vert 0\right\rangle \text{ }|j\in
(-\infty ,N_{s}]\right\} $, The Schr\"{o}dinger equation has the explicit
form%
\begin{eqnarray}
&-Jf_{1}\left( j-1\right) -Jf_{1}\left( j+1\right) =Ef_{1}\left( j\right) ,%
\text{ }\left( j<0\right) &  \label{Seq_g} \\
&-Jf_{1}\left( -1\right) -gf_{1}\left( 1\right) =\left( E+i\gamma \right)
f_{1}\left( 0\right) ,&  \notag \\
&-gf_{1}\left( 0\right) +\sum_{i=1}^{N_{s}}\kappa _{i1}f_{1}\left( i\right)
=Ef_{1}\left( 1\right) &  \notag \\
&\sum_{i=1}^{N_{s}}\kappa _{ij}f_{1}\left( i\right) =Ef_{1}\left( j\right) ,%
\text{ }\left( j\in \left[ 2,N_{s}\right] \right) &  \notag
\end{eqnarray}%
within all the regions. We will show that, such a scattering process can
occur in a Hermitian system.

It has been well known that an imaginary potential, by means of an effective
interaction, can serve as a reduced description for the outer world of an
open system. Along this line, we consider a similar lattice system to $%
H_{\gamma }$, described by a \emph{Hermitian} Hamiltonian $H_{V}$. In this
network, the imaginary potential is replaced by a real potential $V$, and a
semi-infinite chain is added, which acts as the complementary subspace or
outer word if $H_{\gamma }$ is regarded as an open system. The corresponding
Hamiltonian has the form
\begin{eqnarray}
H_{V} &=&H_{lV}+H_{g}+H_{\text{sub}}+H_{\nu }+H_{l\nu }  \label{H_V} \\
H_{lV} &=&-J\overset{-1}{\sum_{l=-\infty }}\left( a_{l}^{\dag }a_{l+1}+\text{%
H.c.}\right) +Va_{0}^{\dag }a_{0}  \notag \\
H_{\nu } &=&-\nu \left( a_{0}^{\dag }b_{1}+b_{1}^{\dag }a_{0}\right)  \notag
\\
H_{l\nu } &=&-J\overset{\infty }{\sum_{l=1}}\left( b_{l}^{\dag }b_{l+1}+%
\text{H.c.}\right) .  \notag
\end{eqnarray}%
where $b_{l}^{\dag }$ is also the creation operator of the boson (or
fermion) at the $l$th site. $H_{lV}$ and $H_{l\gamma }$\ are two
semi-infinite uniform chains with real potential $V$ at the joint point.

A sketch of such a system is given in Fig. 1(b). Note that the two systems
have assigned the same sub-network, $j\in (-\infty ,0]$, which is referred
as the common region. The corresponding scattering wave functions within the
two semi-infinite chains are $f_{2}\left( j\right) a_{j}^{\dag }\left\vert
0\right\rangle $ and $\tilde{f}\left( j\right) b_{j}^{\dag }\left\vert
0\right\rangle $ with%
\begin{eqnarray}
f_{2}\left( j\leq 0\right) &=&e^{ikj}+r_{2}e^{-ikj}  \label{f2} \\
\tilde{f}\left( j>0\right) &=&te^{ikj}.  \notag
\end{eqnarray}

In the basis $\left\{ a_{j}^{\dag }\left\vert 0\right\rangle |j\in (-\infty
,N_{s}],b_{j}^{\dag }\left\vert 0\right\rangle |j\in (1,+\infty ]\right\} $,
the Schrodinger equations are

\begin{eqnarray}
-Jf_{2}\left( j-1\right) -Jf_{2}\left( j+1\right) =Ef_{2}\left( j\right) ,%
\text{ }\left( j<0\right) , &&  \label{Seq_V} \\
-Jf_{2}\left( -1\right) -gf_{2}\left( 1\right) -\nu \tilde{f}\left( 1\right)
=\left( E-V\right) f_{2}\left( 0\right) , &&  \notag \\
-gf_{2}\left( 0\right) +\sum_{i=1}^{N_{s}}\kappa _{i1}f_{2}\left( i\right)
=Ef_{2}\left( 1\right) , &&  \notag \\
\sum_{i=1}^{N_{s}}\kappa _{ij}f_{2}\left( i\right) =Ef_{2}\left( j\right) ,%
\text{ }\left( j\in \left[ 2,N_{s}\right] \right) , &&  \notag
\end{eqnarray}%
and%
\begin{eqnarray}
&-\nu f_{2}\left( 0\right) -J\tilde{f}\left( 2\right) =E\tilde{f}\left(
1\right) ,& \\
&-J\tilde{f}\left( j-1\right) -J\tilde{f}\left( j+1\right) =&E\tilde{f}%
\left( j\right) ,\text{ }\left( j>1\right) .  \notag
\end{eqnarray}%
We can see that the Eqs. (\ref{Seq_g}) and (\ref{Seq_V}) within the common
region have the same form except the potentials at the $0$th site. For the
the same incident plane wave $e^{ikj}$, we have

\begin{equation}
E=-2J\cos k
\end{equation}%
and furthermore, one can find that under the conditions%
\begin{equation}
\nu ^{2}\sin k=\gamma J\text{, }\nu ^{2}\cos k=VJ,  \label{conds}
\end{equation}%
the solutions for $r_{1}$\ and $r_{2}$\ are identical. The above equivalent
conditions can also be given in the energy-dependent form

\begin{equation}
\nu ^{2}=\frac{2\gamma J^{2}}{\Omega }\text{, }V=-\frac{\gamma E}{\Omega }
\label{cond_E}
\end{equation}%
where $\Omega =\sqrt{4J^{2}-E^{2}}$. Then the wave functions (\ref{f1}) and (%
\ref{f2}) within $j\in \left[ -\infty ,N_{s}\right] $ are the same. This
indicates that for the incident plane wave $e^{ikj}$,\ the imaginary
potential can be regarded as a semi-infinite chain for wave escaping. It is
worth to point out that it is a conditional equivalence, which is only for
the specific state. This equivalence is the building block for the
investigation of this paper.

Now we consider the relevant situations derived from the obtained results.
We are interested in the case when the imaginary potential is source like.
Actually, applying time-reversal operation on the above scattering
processes, i.e., taking complex conjugation for the Eqs. (\ref{H_g}) and (%
\ref{H_V}), the corresponding time reversal solutions can be obtained, which
are illustrated in Fig. 1(c, d). The Fig. 1(c) shows that the time reversal
solution is for the new system $H_{\bar{\gamma}}=H_{\gamma }\left( \gamma
\rightarrow -\gamma \right) $. Nevertheless, its counterpart $H_{V}$\ is
invariant under time reversal. The Fig. 1(d) illustrates the corresponding
time reversal process of that in Fig. 1(b). Based on the processes in real
physical systems illustrated in Fig. 1(b, c),\ the physics of the imaginary
potential becomes clear: $-i\gamma $ acts as a drain lead, while $i\gamma $\
acts as a source lead associated with an incoming plane wave. Although this
is not a surprising result, we still verify it explicitly in a strict manner
and will apply it to a $\mathcal{PT}$\textbf{\ }non-Hermitian system.

\section{Resonant transmission condition and $\mathcal{PT}$ symmetry}

The above result is essentially about the stationary state for the infinite
non-Hermitian system. Intuitively, a source and drain could produce a
stationary state in a finite system between them when the gain rate equals
to the loss rate, or in an open system\ with injecting sources and absorbing
sinks. The conceptual framework is required to substantiate this idea. In
this section, we consider the stationary state of a non-Hermitian system
based on the obtained scattering solutions of the imaginary potential.

Although we cannot get the explicit solution about the reflection amplitude $%
r$, the corresponding time-reversal process illustrated in Fig. 1(c)
exhibits the facts: for an incident plane wave with amplitude $1$, the
reflected amplitude from the absorptive potential $-i\gamma $\ is $r$, while
for an incident plane wave with amplitude $r^{\ast }$, the reflected
amplitude from the source potential $i\gamma $\ is just $1$. The fact $%
\left\vert r\right\vert =\left\vert r^{\ast }\right\vert $\ indicates that
if we combine the building blocks Fig. 1(a) and (c) to construct a finite
network with the geometry illustrated in Fig. 2(a), the stationary state may
be formed in the manner: a wave coming from the sources and send back its
time-reversed version. On the other hand, such a configuration has the $%
\mathcal{PT}$\ symmetry spontaneously, which has been shown to process
real-energy eigenstate under certain condition. In the aid of its Hermitian
counterpart shown in Fig. 2(b), one can find that in the case of the
stationary state being formed, it just corresponds to the resonant
transmission. Therefore the existence of the real-energy eigenstate of a $%
\mathcal{PT}$\ symmetric system is connected to the occurrence of resonant
transmission in a Hermitian system. It follows that we can find an
alternative Hermitian counterpart to a $\mathcal{PT}$\ symmetric Hamiltonian
in the sense that they share the same eigenfunction within the common
region. This should be the way more directly associated with the physics of
the $\mathcal{PT}$\ symmetric system. In the following sections, we will
study the formation of the $\mathcal{PT}$ symmetrical state by dealing with
the more tractable models.

%%%%%%%%%%%%%%%%%%%%%%%%%%%%%%%%%%%%%%%%%%%%%%%%%%%%%%%%%%%%%%%%%%%%%%%%
\begin{figure}[tbp]
\includegraphics[ bb=86 316 525 755, width=5.5 cm, clip]{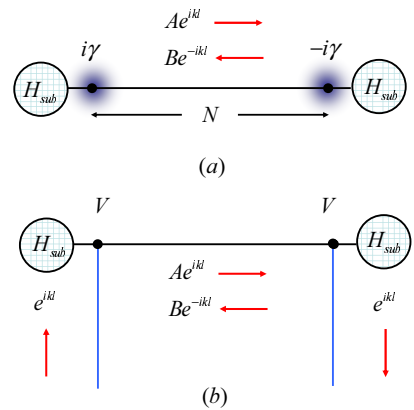}
\caption{(Color online) Schematic illustration of configurations of a
non-Hermitian tight-binging network with $\mathcal{PT}$ symmetry and its
Hermitian counterpart. They are constructed based on the building blocks
represented in Fig. 1. It indicates that an eigenstate (stationary state) of
(a) corresponds to a resonant transmission state of (b).}
\end{figure}

%%%%%%%%%%%%%%%%%%%%%%%%%%%%%%%%%%%%%%%%%%%%%%%%%%%%%%%%%%%%%%%%%%%%%%%%%

%%%%%%%%%%%%%%%%%%%%%%%%%%%%%%%%%%%%%%%%%%%%%%%%%%%%%%%%%%%%%%%%%%%%%%%%
\begin{figure}[tbp]
\includegraphics[ bb=110 317 505 745, width=5.5 cm, clip]{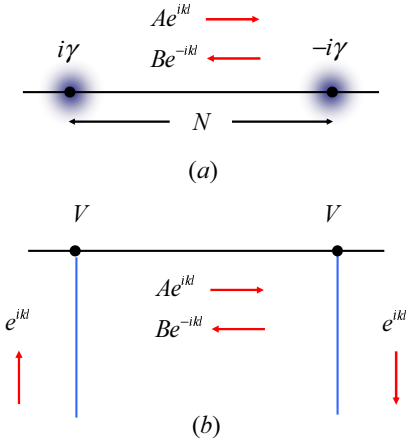}
\caption{(Color online) Schematic illustrations of tight-binging network
with $\mathcal{PT}$\ symmetry and its Hermitian counterpart, where $H_{\text{%
sub}}$\ is a simple chain of length $N_{s}$. Exact Bethe ansatz solution
shows that the eigenstate of (a) accords to the equal-energy resonant
scattering state of (b) under the condition (\protect\ref{conds}).}
\label{fig3}
\end{figure}

%%%%%%%%%%%%%%%%%%%%%%%%%%%%%%%%%%%%%%%%%%%%%%%%%%%%%%%%%%%%%%%%%%%%%%%%%

\section{Illustrative examples}

In this section, we investigate a simple exactly solvable system to
illustrate the main idea of this paper. In order to exemplify the above
mentioned analysis of relating the stationary states of a non-Hermitian $%
\mathcal{PT}$\ symmetric Hamiltonian and a Hermitian one, we take $H_{\text{%
sub}}$\ to be the simplest network: a uniform chain. Then the sample
Hamiltonian has the form%
\begin{eqnarray}
H_{\gamma \bar{\gamma}} &=&-J\overset{N+2N_{s}-1}{\sum_{l=1}}\left(
a_{l}^{\dag }a_{l+1}+\text{H.c.}\right)  \label{H_gg} \\
&&+i\gamma a_{N_{s}+1}^{\dag }a_{N_{s}+1}-i\gamma a_{N+N_{s}}^{\dag
}a_{N+N_{s}},  \notag
\end{eqnarray}%
which is sketched in Fig. 3(a). It has $\mathcal{PT}$ symmetry, i.e., $%
H_{\gamma \bar{\gamma}}^{\mathcal{PT}}=\mathcal{PT}H_{\gamma \bar{\gamma}}%
\mathcal{PT}=H_{\gamma \bar{\gamma}}$, which has been studied systematically
in the\ case with zero $N_{s}$ \cite{JLPT}. Here $\mathcal{P}$ and $\mathcal{%
T}$\ represent the space-reflection operator (or parity operator) and the
time-reversal operator respectively. The corresponding Hermitian Hamiltonian
reads%
\begin{eqnarray}
H_{VV} &=&\left[ -J\sum_{l=1}^{N+2N_{s}-1}a_{l}^{\dag }a_{l+1}-J\overset{\pm
\infty }{\sum_{l=\pm 1}}b_{l}^{\dag }b_{l\pm 1}\right.  \label{H_VV} \\
&&\left. -\nu \left( a_{N_{s}+1}^{\dag }b_{-1}+a_{N_{s}+N}^{\dag
}b_{1}\right) +\text{H.c}\right]  \notag \\
&&+V\left( a_{N_{s}+1}^{\dag }a_{N_{s}+1}+a_{N_{s}+N}^{\dag
}a_{N_{s}+N}\right) ,  \notag
\end{eqnarray}%
which is sketched in Fig. 3(b). It is a $\mathcal{P}$\ symmetric system,
i.e., $\left[ \mathcal{P},H\right] =0$, which has been studied in the\
framework of Bethe ansatz for case with zero $V$ \cite{JLAB}.\ The effects
of $\mathcal{P}$ and $\mathcal{T}$\ on a discrete system are

\begin{equation}
\mathcal{T}i\mathcal{T}=-i\text{, }\mathcal{P}a_{l}^{\dag }\mathcal{P}%
=a_{N+2N_{s}+2-l}^{\dag }\text{, }\mathcal{P}b_{l}^{\dag }\mathcal{P}%
=b_{-l}^{\dag }.
\end{equation}%
Note the region $\left\{ a_{l}^{\dag }\left\vert 0\right\rangle ,\text{ }%
l\in \left( 1,N+2N_{s}\right) \right\} $ is regarded as the common region of
the two models. In the following, we present the analytical results in the
framework of Bethe ansatz for the two models in order to perform a
comprehensive study.

\subsection{$\mathcal{PT}$ chain $H_{\protect\gamma \bar{\protect\gamma}}$}

According to the $\mathcal{PT}$-symmetric\ quantum mechanics \cite{Bender 02}%
, system $H_{\gamma \bar{\gamma}}$ can be further classified to be either\
unbroken $\mathcal{PT}$ symmetry or broken $\mathcal{PT}$ symmetry, which
depends on the symmetry of the eigenstates $\left\vert \psi _{k}^{\gamma
}\right\rangle $ in different regions of $\gamma $. The time-independent Schr%
\"{o}dinger equation is%
\begin{equation}
H_{\gamma \bar{\gamma}}\left\vert \psi _{k}^{\gamma }\right\rangle
=\varepsilon _{k}^{\gamma }\left\vert \psi _{k}^{\gamma }\right\rangle
\end{equation}%
with corresponding eigenvalue $\varepsilon _{k}^{\gamma }$. The system is
unbroken $\mathcal{PT}$ symmetry if \textit{all} the eigenfunctions have $%
\mathcal{PT}$ symmetry%
\begin{equation}
\mathcal{PT}\left\vert \psi _{k}^{\gamma }\right\rangle =\left\vert \psi
_{k}^{\gamma }\right\rangle  \label{PT state}
\end{equation}%
and all the corresponding eigenvalues are real simultaneously. This
classification depends on the value of the parameter $\gamma $. Beyond the
unbroken $\mathcal{PT}$ symmetric region the system is broken $\mathcal{PT}$
symmetry, where Eq. (\ref{PT state}) does not hold for \textit{all} the
eigenfunctions and the eigenvalues of broken $\mathcal{PT}$ symmetric
eigenfunctions are complex. We denote the single-particle eigenfunction in
the form $\left\vert \psi _{k}^{\gamma }\right\rangle =\sum f_{k}^{\gamma
}\left( l\right) a_{l}^{\dag }\left\vert 0\right\rangle $.

According to Bethe ansatz method, the eigenstates can be in the form of

\begin{equation}
f_{k}^{\gamma }=\left\{
\begin{array}{ll}
C_{L}e^{ikj}+D_{L}e^{-ikj}, & j\in \left[ 1,N_{s}+1\right] \\
Ae^{ikj}+Be^{-ikj}, & \left[ N_{s}+1,N_{s}+N\right] \\
C_{R}e^{ikj}+D_{R}e^{-ikj}, & \left[ N_{s}+N,N+2N_{s}\right]%
\end{array}%
\right. .  \label{f_gamma}
\end{equation}%
The coefficients $\left\{ A,B,C_{L(R)},D_{L(R)}\right\} $\ and the quasi
momentum $k$ are to be determined by matching conditions%
\begin{eqnarray}
&&f_{k}^{\gamma }\left( j+0^{+}\right) =f_{k}^{\gamma }\left( j+0^{-}\right)
,  \label{continuity} \\
&&\left( j=N_{s}+1,N_{s}+N\right)  \notag
\end{eqnarray}%
and the corresponding Schr\"{o}dinger equations of $j$ ($j\neq N_{s}+1$, $%
N+N_{s})$ in the center of the system,
\begin{equation}
-Jf_{k}^{\gamma }\left( j+1\right) -Jf_{k}^{\gamma }\left( j-1\right)
=\varepsilon _{k}^{\gamma }f_{k}^{\gamma }\left( j\right) ,  \label{Seq1}
\end{equation}%
the Schr\"{o}dinger equations of $j$ for the sites $j=N_{s}+1$ and $N_{s}+N$%
,
\begin{eqnarray}
&&-Jf_{k}^{\gamma }\left( N_{s}+2\right) -Jf_{k}^{\gamma }\left( N_{s}\right)
\label{Seq2} \\
&=&\left( \varepsilon _{k}^{\gamma }-i\gamma \right) f_{k}^{\gamma }\left(
N_{s}+1\right) ,  \notag \\
&&-Jf_{k}^{\gamma }\left( N_{s}+N+1\right) -Jf_{k}^{\gamma }\left(
N_{s}+N-1\right)  \notag \\
&=&\left( \varepsilon _{k}^{\gamma }+i\gamma \right) f_{k}^{\gamma }\left(
N_{s}+N\right) ,  \notag
\end{eqnarray}%
and for the edges of the system $j=1$ and $N+2N_{s}$,
\begin{eqnarray}
&-Jf_{k}^{\gamma }\left( 2\right) =\varepsilon _{k}^{\gamma }f_{k}^{\gamma
}\left( 1\right) & \\
&-Jf_{k}^{\gamma }\left( N+2N_{s}-1\right) =\varepsilon _{k}^{\gamma
}f_{k}^{\gamma }\left( N+2N_{s}\right) .&  \notag
\end{eqnarray}

These lead to the equation of $k$%
\begin{eqnarray}
&&-\gamma ^{2}\chi _{k}^{2}\left[ e^{ik\left( N-1\right) }-e^{-ik\left(
N-1\right) }\right]  \label{e} \\
&=&J^{2}\left[ e^{ik\left( N+2N_{s}+1\right) }-e^{-ik\left(
N+2N_{s}+1\right) }\right] ,  \notag
\end{eqnarray}%
where%
\begin{equation}
\chi _{k}=\frac{e^{ik\left( N_{s}+1\right) }-e^{-ik\left( N_{s}+1\right) }}{%
e^{ik}-e^{-ik}}.
\end{equation}%
The solutions of $k$ can be classified in two categories: \textit{physical}
and \textit{unphysical} states, which correspond to real and complex $k$,
respectively. The corresponding energy are real or complex and in the form%
\begin{equation}
\varepsilon _{k}^{\gamma }=-J\left( e^{ik}+e^{-ik}\right) .
\end{equation}%
A straightforward algebra shows that there are at least $\left( N-1\right) $%
\ solutions of real $k$ for arbitrary $\gamma /J$. In this paper, we only
focus on the physics counterparts of these states rather than the detailed
form of the solutions.

\subsection{Hermitian counterpart $H_{VV}$}

For the Hamiltonian Eq. (\ref{H_VV}), it has $\mathcal{P}$ and $\mathcal{T}$%
\ symmetry simultaneously, i.e., $\mathcal{P}H_{VV}\mathcal{P}=\mathcal{T}%
H_{VV}\mathcal{T}=H_{VV}$. Nevertheless, for a scattering state, a plane
wave comes from the leftmost, the $\mathcal{P}$\ and $\mathcal{T}$\ symmetry
are broken. We will show that under certain conditions, the wave function
within the common region of $H_{\gamma \bar{\gamma}}$\ and $H_{VV}$\ has $%
\mathcal{PT}$\ symmetry.

We can set the scattering wave function in the form of

\begin{equation}
f_{k}^{V}=\left\{
\begin{array}{ll}
C_{L}^{s}e^{ikj}+D_{L}^{s}e^{-ikj}, & \left[ 1,N_{s}+1\right] \\
A^{s}e^{ikj}+B^{s}e^{-ikj}, & \left[ N_{s}+1,N_{s}+N\right] \\
C_{R}^{s}e^{ikj}+D_{R}^{s}re^{-ikj}, & \left[ N_{s}+N,N+2N_{s}\right]%
\end{array}%
\right. ,
\end{equation}%
and%
\begin{equation}
\tilde{f}_{k}^{V}=\left\{
\begin{array}{ll}
e^{ikj}+re^{-ikj} & j\in (-\infty ,-1] \\
te^{ikj}, & [1,+\infty )%
\end{array}%
\right. ,
\end{equation}%
where $f_{k}^{V}$\ represents the one within the common region, while $%
\tilde{f}_{k}^{V}$\ represents the one\ in the two leads. The reflection and
transmission amplitudes $r$, $t$,\ coefficients $\left\{
A^{s},B^{s},C_{L(R)}^{s},D_{L(R)}^{s}\right\} $\ and the quasi momentum $k$
are to be determined by matching conditions
\begin{eqnarray}
f_{k}^{V}\left( j+0^{+}\right) =f_{k}^{V}\left( j+0^{-}\right) , && \\
\left( j=N_{s}+1,N+N_{s}\right) &&  \notag \\
\tilde{f}_{k}^{V}\left( N_{s}+1\right) =f_{k}^{V}\left( N_{s}+1\right) &&
\notag \\
\tilde{f}_{k}^{V}\left( N_{s}+N\right) =f_{k}^{V}\left( N_{s}+N\right) &&
\notag
\end{eqnarray}%
and the corresponding Schr\"{o}dinger equations of $j$ in the center of the
system,%
\begin{eqnarray}
-Jf_{k}^{V}\left( j+1\right) -Jf_{k}^{V}\left( j-1\right) &=&\varepsilon
_{k}^{\gamma }f_{k}^{V}\left( j\right) \\
-J\tilde{f}_{k}^{V}\left( j+1\right) -J\tilde{f}_{k}^{V}\left( j-1\right)
&=&\varepsilon _{k}^{\gamma }\tilde{f}_{k}^{V}\left( j\right)  \notag
\end{eqnarray}%
Schr\"{o}dinger equations of $j$ for the connection sites $N_{s}+1,N_{s}+N$,%
\begin{eqnarray}
&&-\nu \tilde{f}_{k}^{V}\left( -1\right) -Jf_{k}^{V}\left( N_{s}+2\right) \\
&&-Jf_{k}^{V}\left( N_{s}\right) =\left( \varepsilon _{k}^{\gamma }-V\right)
f_{k}^{V}\left( N_{s}+1\right) ,  \notag \\
&&-\nu \tilde{f}_{k}^{V}\left( 1\right) -Jf_{k}^{V}\left( N_{s}+N+1\right)
\notag \\
&&-Jf_{k}^{V}\left( N_{s}+N-1\right) =\left( \varepsilon _{k}^{\gamma
}-V\right) f_{k}^{V}\left( N_{s}+N\right) ,  \notag
\end{eqnarray}%
and for the edges of the system $j=1$, $N+2N_{s}$,%
\begin{eqnarray}
-Jf_{k}^{V}\left( 2\right) &=&\varepsilon _{k}^{\gamma }f_{k}^{V}\left(
1\right) \\
-Jf_{k}^{V}\left( N+2N_{s}-1\right) &=&\varepsilon _{k}^{\gamma
}f_{k}^{V}\left( N+2N_{s}\right) .  \notag
\end{eqnarray}%
The solution for reflection amplitude $r$ is given, after a straightforward
algebra, by

\begin{equation}
r=\frac{2i\nu ^{2}\xi \sin \left( k\right) \sin ^{2}\left[ k\left(
N-1\right) \right] }{J^{2}\sin ^{2}\left( k\right) -J^{2}\xi ^{2}\sin ^{2}%
\left[ k\left( N-1\right) \right] }-1,  \label{r}
\end{equation}%
where%
\begin{equation}
\xi =\frac{V}{J}+\frac{\sin \left[ kN\right] }{\sin \left[ k\left(
N-1\right) \right] }-\frac{\nu ^{2}}{J^{2}}e^{ik}-\frac{\sin \left[ kN_{s}%
\right] }{\sin \left[ k\left( N_{s}+1\right) \right] }.  \label{eta}
\end{equation}%
We are interested in the resonant transmission state, which should be
relevant to the $\mathcal{PT}$ eigenstate of $H_{\gamma \bar{\gamma}}$,
according to the above time reversal analysis. For $r=0$ we have

\begin{eqnarray}
&&2i\nu ^{2}\xi \sin \left( k\right) \sin ^{2}\left[ k\left( N-1\right) %
\right]  \label{resonance} \\
&=&J^{2}\sin ^{2}\left( k\right) -J^{2}\xi ^{2}\sin ^{2}\left[ k\left(
N-1\right) \right] .  \notag
\end{eqnarray}

The analysis in Sec. II indicates that, under the conditions (\ref{conds}),
there should be a resonant transmission state corresponding to the
eigenstate of (\ref{H_gg}). In fact, substituting (\ref{conds}) into (\ref%
{resonance}), one can obtain the%
\begin{eqnarray}
&&\gamma ^{2}\sin ^{2}\left[ k\left( N_{s}+1\right) \right] \sin \left[
k\left( 1-N\right) \right]  \label{sin} \\
&=&J^{2}\sin ^{2}\left( k\right) \sin \left[ k\left( N+2N_{s}+1\right) %
\right] ,  \notag
\end{eqnarray}%
which is just the reduced form of the Eq. (\ref{e})\ for the \textit{real}
quasi momentum $k$. It can be shown exactly that there are at least $N-1$\
real solutions for the Eq. (\ref{sin}). This exhibits the connection between
the two models (\ref{H_gg}) and (\ref{H_VV}) predicted by the above
mentioned analysis. Accordingly, we have

\begin{equation}
\frac{C_{L(R)}}{C_{L(R)}^{s}}=\frac{D_{L(R)}}{D_{L(R)}^{s}}=\frac{A}{A^{s}}=%
\frac{B}{B^{s}}
\end{equation}%
which shows that both functions within the common region are identical,
i.e., the scattering wave function $\left\{ r,t\text{, }A^{s},B^{s}\text{, }%
C_{L(R)}^{s},D_{L(R)}^{s}\right\} $\ is the analytical continuation of the
one $\left\{ A,B,C_{L(R)},D_{L(R)}\right\} $.

Thus two eigenstates of the Hamiltonian (\ref{H_gg})\ belong to the resonant
transmission scattering states of \textit{two} \textit{different} systems
(with different $V$ and $v$). In this sense, nonorthogonality of the
eigenstates of a Pseudo-Hermitian Hamiltonian is obvious. In general the
norm of a wave function is conserved in a Hermitian system, while the norm
of many components of the wave function is not. These wave components may be
associated with a truncated basis set, or a subspace of the full Hilbert
space. The common region of $H_{\gamma \bar{\gamma}}$\ and $H_{VV}$\ is a
concrete example in the discrete system to demonstrate this fact.

\section{Conclusion and discussion}

In conclusion, we have presented an alternative way of finding the link
between a $\mathcal{PT}$\ non-Hermitian Hamiltonian and a Hermitian one,
based on the analysis of the scattering problem for an imaginary potential.
We have found that the on-site absorptive imaginary potential can be
equivalent to an attached semi-infinite chain as a drain with respect to a
specific wave scattering problem. Applying this result and its extension to
the time reversal process on the $\mathcal{PT}$\ non-Hermitian system, the
eigenstate problem is connected to that of resonant transmission problem in
the corresponding Hermitian system. It is shown that any real-energy
eigenstate of a $\mathcal{PT}$ tight-binding lattice with on-site imaginary
potentials shares the same wave function with a resonant transmission state
of the corresponding Hermitian lattice embedded in a chain.

It is not surprising that the $\mathcal{PT}$ eigenstate has connection to
the resonance transmission state of the extended system. In general, for an
infinite system with parity ($\mathcal{P}$) symmetry, it also has the time ($%
\mathcal{T}$) reversal symmetry. An arbitrary scattering state, as an
eigenstate of the system, probably breaks the $\mathcal{P}$, $\mathcal{T}$\
and also $\mathcal{PT}$\ symmetries simultaneously. Interestingly, in the
case of the resonance transmission, the corresponding wave function
possesses $\mathcal{PT}$\ symmetry. Such wave functions can be the
eigenstates of the $\mathcal{PT}$\ non-Hermitian system with the real
eigenvalues.

We acknowledge the support of the CNSF (Grants No. 10874091 and No.
2006CB921205).

\end{document}